\documentclass{article}
\begin{document}

\begin{flushleft}
KL TH 00/08
\end{flushleft}
\vglue.1in

\begin{center}
\textbf{Why do we need supersymmetry?}
\vglue.2in

Sergei V. Ketov 

{\it Department of Theoretical Physics\\
     Erwin Schr\"odinger Strasse \\
     University of Kaiserslautern}\\
{\it 67653 Kaiserslautern, Germany}
\vglue.1in
{\sl ketov@physik.uni-kl.de}
\end{center}

The word supersymmetry came to me for the first time in 1980 when I was the 
third-year student at the Physics Department of Tomsk State University in 
Western Siberia, in the former Soviet Union. Unlike the physics students in 
Moscow (not to mention those in the West), we were quite isolated from the Big
 Science, since there was no single laboratory to be related with either
experimental or theoretical high energy physics in Tomsk and, of course, there
were no local traditions at all. The first breakthrough in Tomsk came a few 
years earlier, with the opening of the quantum field theory group under the 
supervision of Prof. V. Bagrov, an expert in exact classical solutions to the 
Dirac equation in external electromagnetic fields, who was lucky to spent some
 time at the Physics Department of Moscow State University. Learning about the
 Dirac equation from the lectures of Prof. V. Bagrov appeared to be my first 
step towards supersymmetry. 

In 1980 Prof. E. Fradkin from the Lebedev Physical Institute in Moscow asked 
Prof. V. Bagrov to send some of the best physics students from Tomsk to 
Moscow, for doing research under his supervision. The only condition of 
Prof. E. Fradkin was that the students should be maximally 20 years old, 
since, otherwise, it would be too late for them to study theoretical physics 
(!) I was amongst these students. Of course, we were not taken seriously in
Moscow, and all of us were very embarrassed there, since we didn't understand 
a word during the first meeting with Prof. E. Fradkin. One should also mention
 that the way of dealing with students (at least with us) was very cruel in 
Moscow: our senior supervisors expected from us to know everything, 
from quantization of non-abelian field theories until N=8 supergravity, as the
 pre-requisite for any serious discussion, which was, of course, unfair. From 
our first trip to Moscow we just learned a few foreign words, with 
`supersymmetry' being one of them. The first tough lesson in Moscow gave us 
the first motivation to learn supersymmetry in Tomsk, simply because there was
 no other challenging message around. Perhaps, it is hard to imagine now, 
what it was to learn supersymmetry in Siberia, in the absense 
of regular western literature and world-wide-web (not to mention a personal
computer). We were getting the scientific papers privately from the capital, 
which implied regular travel about 3100 km from Tomsk to Moscow and back, 
spending a lot of time in the Moscow libraries, and copying hundreds of pages 
there (no xerox machines were available in Tomsk).  

Fortunately, there were a few senior people in Tomsk at that time, who helped 
us with our education in quantum field theory: for example, Prof. I. Tyutin 
(T in the BRST) from the Lebedev Institute. We were suddenly offered  a plenty
 of lectures and seminars about group theory, field theory, supersymmetry 
and quantization (everything on the top of regular courses at Tomsk 
University), much earlier (and, sometimes, instead) of many standard courses in
 physics. Because of this background, when I became a graduate student 
and again came to the Lebedev Institute in Moscow, I had no doubt that 
supersymmetry is the only thing worthy to be studied.

Now, after 20 years, I ask myself again, (i) how should we classify the subject
of supersymmetry, (ii) why do we need supersymmetry, and (iii) what is the 
future of supersymmetry ? I would like to offer my own, very personal view on 
these matters.  

As is well known, the first papers about supersymmetry appeared in the early
seventies, in the former Soviet Union. Drs. Yu. Gol'fand and E. Lichtman from
the Lebedev Physical Institute in Moscow found a supersymmetric extension of
 the Poincar\'e algebra for the first time, whereas Prof. D. Volkov and
Dr. V. Akulov from the Phys.-Technical Institute in Char'kov (Ukraine) 
discovered a field-theoretical model with spontaneousy broken supersymetry.
However, these fundamental discoveries were not immediately recognized or
appreciated by the very strong community of theoretical physicists in the 
former Soviet Union and, especially, in and around Moscow. Only after the 
fundamental papers of Prof. B. Zumino and Prof. J. Wess from CERN, 
who pioneered the representation theory of supersymmetry in field theory, 
the explosion of papers devoted to supersymmetry really began. 
The natural question arises why
the discovery of supersymmetry was largely ignored in the former Soviet Union
until the Wess-Zumino contributions? 
I believe that the reason was two-fold. On the
one side, the early inventors of supersymmetry apparently didn't appreciate 
themselves the true meaning of their discoveries. For instance, the 
Gol'fand-Lichtman paper was merely devoted to presenting a new super-algebra, 
 whereas the Akulov-Volkov investigation was motivated by the search for a
non-linear Lagrangian describing neutrino as Goldstone fermion, without looking
 for linear realizations of supersymmetry and its relation to spacetime 
symmetries. On the other side, the message came from the researchers who 
didn't  belong to the top brass of the (highly hierarchical) scientific 
establishment in Moscow. Being a student at the Lebedev Institute in Moscow, 
I got an impression that, for example, Yu. Gol'fand was often treated as 
a `crazy guy' amongst his colleagues. I attended two of his seminars, and
I can now acknowledge this opinion. Perhaps, one ought to be crazy in 
order to generate a crazy idea which is crazy enough to be right! On the
contrary, the Theory Department of CERN was very quick in recognizing and 
appreciating the fundamenetal meaning of supersymmetry as the unifying 
symmetry between bosons and fermions (i.e. the right place, the right people 
and the right time). 
 
The fact that supersymmetry was never experimentally observed or confirmed 
does not apparently bother most theoretical physicists at all. After all, the
theoretical fundament of supersymmetry is much broader and solider than that
of many other modern theoretical constructions. As a result, the need to
motivate supersymmetry itself totally disappeared from the current literature
dealing with supersymmetry. Theoretical consistency and experts opinion
have long substituted the objective experimental criteria in the modern
theoretical high energy physics, including supersymmetry. Moreover, it is
sometimes very difficult, if not possible, to distinguish between proved
statements and conjectures either in the current literature or in the 
hep-th archive. From this perspective, supersymmetry can be considered as a 
kind of art, or as part of mathematics. I would, nevertheless, refrain from 
identifying supersymmetry with the intellectual entertainment for qualified 
scientists.  

Supersymmetry is not only the part of theory. It also creates jobs and 
attracts money. Any new (bosonic) field theory entering the theory market may 
be supersymmetrized; this gives the unlimited source of motivation for 
writing new theoretical papers and Ph.D. Theseses, as well as demanding new 
postdoc positions from the funding agencies. Once the abstract theory language
 of supersymmerty had become available in physical terms for experimental 
physicists, the search for supersymmetry turned into one of the main 
topics in their agenda, with all its cosequences to be related with a 
construction of new expensive experimental devices like LHC. Hence, 
supersymmetry is the business enterprise also.

Yet another unusual view on supersymmetry is provided by evaluating the 
problems in supersymmetry as the challenge for supersymmetry experts. 
For example, once a bosonic theory is supersymmetrized once, it may be 
supersymmetrized twice, etc. with the increasing level of complexity. One may
also go in the opposite direction: once a model with partial (1/2) 
supersymmetry breaking is found, one may try to get other patterns with 1/4, 
1/8 or even 3/16 of supersymmetry breaking, which are definitely much harder to
construct. The challenge results in a competition, the competition gives rise 
to winners and losers, the winners get recognition and prices. Hence, it 
is also possible to identify supersymmetry with a kind of sport too.

If something can be simultaneously interpreted as science, business, art and 
sport, it is definitely the important subject that is going to stay with us
forever. In the rest of this paper I would like to concentrate on the 
functional role of supersymmetry in modern theoretical high-energy physics.

The standard motivation for supersymmetry is based on its interpretation as the
unification symmetry between the fundamental bosonic and fermionic degrees of
freedom. Supersymmetry is also known to be the only non-trivial way of 
unifying the spacetime symmetries and the internal symmetries. Being the 
`square root' of spacetime, local supersymmerty immediately implies gravity. 
This motivation was put forward in the early days of supersymmetry and 
supergravity towards a formulation of the unified field theory of all 
fundamental physical interactions, including gravity. The hope was that the 
maximal supersymmetry (realized in N=8 supergravity) could automatically care
 of the problems beyond the Standard Model. This didn't happen, and it gives us
the lesson that a relation between supersymmetry and particle physics is less 
straightforward as it seemed in the beginning of the supersymmetry era.
 
The development of supersymmetry is to be compared with the (apparently 
unrelated) development of the dual models (now known as string theory), before
their unification proposed by Prof. J. Schwarz from Caltech and his 
collaborators. In fact, the fermionic dual models already had (what is now 
called) world-sheet supersymmetry, so that it was not very surprising that 
the Wess-Zumino work appeared to be a catalyzator for a discovery of 
spacetime supersymmetry in the fermionic dual models (now called superstring 
theory). String theory also gives us another lesson that the naive increase in
the amount of supersymmetry is not always productive: for example, the 
world-sheet supersymmetry of the NSR string model was abandomed in favor of 
spacetime supersymmetry, the N=2 world-sheet supersymmetric strings are 
inconsistent at the one-loop (string) level, while the strings with N=4 
world-sheet supersymmetry do not have the spacetime interpretation at all
(their critical dimension is negative or zero). 

A deeper consequence of supersymmetry is cancellation amongst Feynman graphs
(and their ultra-violet divergences) between bosonic and fermionic 
contributions. This is not only crucial for particle physics (e.g. as regards
the hierarchy problem), but is of paramount importance for getting solutions 
to quantum gauge theories and strings. As is well-known, the description of the
non-abelian quantum gauge theories in terms of the fundamental (Yang-Mills) 
variables becomes invalid below some energy scale, due to singularities in 
quantum perturbation theory. As a result, the strong coupling description in 
these theories (like QCD) is out of reach. The main physical obstruction is the
complicated vacuum structure of the bosonic gauge theories, which results in
the (theoretically) uncontrollable screening of charges, etc. Supersymmetry
causes the cancellation between screening and anti-screening of the bosonic
and (very specific) fermionic contributions, which greatly simplifies the
low-energy behaviour in the supersymmetric quantum gauge field theories. If
the amount of supersymmetry is enough (as it happens to be the case
in the N=2 supersymmetric quantum gauge field theories in four spacetime 
dimensions), the exact low-energy solutions are possible, as was demonstrated 
in the seminal papers of Prof. N. Seiberg and Prof. E. Witten from Princeton 
in 1994. Without supersymmetry, instanton contributions are plagued by
infra-red divergences.

We can, therefore, conclude that there is the conflict between the `realistic'
 (phenomenological and nonsupersymmetric) field theories, which are 
best exemplified by the non-solvable Standard Model, and the supersymmetric 
gauge theories which may be solvable but are certainly non-realistic. This
conflict reminds me the conflict between the (unrealistic) Yang-Mills 
theories and their (realistic) spontaneously broken counterparts, which is
resolved by the Higgs effect. One expects that the ultimate marriage
of supersymmetry and phenomenology can only happen after a super-Higgs
effect of spontaneous supersymmetry breaking. Spontaneous breaking of any
symmetry allows us to keep control over the effective action. Spontaneous
breaking of supersymmetry naturally implies the existence of the corresponding
Goldstone action whose structure is uniquely determined by the broken
supersymmetry. This mechanism is realized in the D-branes and M-theory, which
has the promise to be the ultimate unified theory of Nature. Supersymmetry 
then plays the role of the universal regulator which puts strong coupling under
control and eliminates unphysical degrees of freedom (like a tachyon). This
may imply an even grater role of supersymmetry in making the supersymmetric 
non-abelian quantum gauge field theories and superstring theory to be well 
defined beyond quantum perturbation theory.  

Yet another example in support of the last conjecture is provided by the
AdS/CFT correspondence.  As is widely believed, the QCD confinement is a 
non-perturbative solution to a four-dimensional quantum $SU(N_{\rm c})$  
gauge field theory with $N_{\rm c}=3$. A formal proof of the colour 
confinement amounts to a derivation of the area law for a  Wilson loop $W[C]$.
The so-called `string' Ansatz 
$$ W[C] \sim \int_{{\rm surfaces~}\Sigma,\atop \partial\Sigma=C}\; 
\exp\left(-S_{\rm string}\right) $$
clearly shows that the effective degrees of freedom (or collective 
coordinates) in QCD at strong coupling (in the infrared) are the (QCD) strings
whose world-sheets are given by the surfaces $\Sigma$, and whose dynamics is 
governed by a string action $S_{\rm string}$. The fundamental 
(Schwinger-Dyson) equations of QCD can be put into the equivalent form of the
(infinite chain) equations for the Wilson loop. This chain of loop equations 
drastically simplifies at large number of colours $N_{\rm c}$ to a 
single closed equation known as the  Makeenko-Migdal (MM) loop equation. Only
planar Feynman graphs survive in this limit. Unfortunately, such approach was
never successful in the past, largely because it was unable to take into 
account quantum renormalization and fix the relevant string action 
$S_{\rm string}$. The first problem may be circumvented via replacing QCD by 
the N=4 Supersymmetric Yang-Mills (SYM) theory that is known to be UV-finite
 and conformally invariant. As was conjectured by Maldacena, the N=4 SYM 
theory is dual to the IIB superstring theory in the ${\rm AdS}_5\times S^5$ 
background. The Maldacena conjecture can therefore be interpreted as the 
particular Ansatz for the string action, 
$S_{\rm string}=S_{{\rm IIB/AdS}_5\times S^5}$, as regards a solution to the 
N=4 supersymmetric (MM) loop equation, provided that spacetime is identified 
with the boundary of the  Anti-de-Sitter space ${\rm AdS}_5$. This CFT/AdS 
correspondence gives rise to simple mechanisms for simulating confinement and
generating the mass gap after breaking the conformal invariance and 
supersymmetry in the `finite-temperature' versions of Anti-de Sitter spaces,
as was demonstrated by Prof. E. Witten in 1998. These recent results lend 
further support for the role of supersymmetry as the universal regulator in
quantum field theory and strings, which seems to be indispensable for their
non-perturbative definition. 

Anyway, supersymmetry is fun, and it is certainly going to be with us in any
forseeable future.

The number of relevant papers about supersymmetry is very large, while they 
can be easily identified  when using the standard databases in theoretical 
high-energy physics, available in internet. So I decided to skip all 
references. 

The idea to write down these notes came to me in response to the question 
put in the title that was raised by a student during my lecture.

\end{document}